# Thermodynamic phase diagram of static charge order in underdoped YBa$_2$Cu$_3$O$_y$

David LeBoeuf[1], S. Krämer[2], W.N. Hardy[3,4], Ruixing Liang[3,4], D.A. Bonn[3,4], and Cyril Proust[1,4]

[1] *Laboratoire National des Champs Magnétiques Intenses, UPR 3228, (CNRS-INSA-UJF-UPS), Toulouse 31400, France.*

[2] *Laboratoire National des Champs Magnétiques Intenses, UPR 3228, (CNRS-INSA-UJF-UPS), Grenoble France.*

[3] *Department of Physics and Astronomy, University of British Columbia, Vancouver V6T 1Z1, Canada.*

[4] *Canadian Institute for Advanced Research, Toronto M5G 1Z8, Canada.*

The interplay between superconductivity and any other competing order is an essential part of the long-standing debate on the origin of high temperature superconductivity in cuprates[1,2]. Akin to the situation of heavy fermions, organic superconductors and pnictides, it has been proposed that the pairing mechanism in cuprates comes from fluctuations of a nearby quantum phase transition[3]. Recent evidence of charge modulation[4] and the associated fluctuations[5,6,7] in the pseudogap phase of YBa$_2$Cu$_3$O$_y$ make charge order a likely candidate for a competing order. However, a thermodynamic signature of the charge ordering phase transition is still lacking. Moreover, whether such charge order is one- or two-dimensional is still controversial but pivotal for the understanding the topology of the reconstructed Fermi surface[8,9]. Here we address both issues by **measuring sound velocities in YBa$_2$Cu$_3$O$_{6.55}$ in high magnetic fields, a powerful thermodynamic probe to detect phase transitions. We provide the first thermodynamic signature of the field-induced charge ordering phase transition in YBa$_2$Cu$_3$O$_y$ allowing construction of a field-temperature phase diagram, which reveals the competing nature of this charge order. The comparison of different acoustic modes indicates that the charge modulation has a two-dimensional character, thus imposing strong constraints on Fermi surface reconstruction scenarios.**

In most La-based cuprate superconductors, static order of both spin and charge (so-called stripe order) has been unambiguously identified by spectroscopic and thermodynamic probes[1,2]. At low temperature, magnetic fields weaken superconductivity and at the same time reinforce the magnitude of such orders[11,12]. Since the superconducting transition T$_c$ in the La-based materials is substantially lower than in other cuprate materials, it has been argued that stripe order is detrimental to high temperature superconductivity. On the other hand in underdoped YBa$_2$Cu$_3$O$_y$ (YBCO), there are also indications for the existence of a stripe order[13] even though $T_c$ = 94 K at optimal doping. The discovery of quantum oscillations[14] combined with the negative Hall[15] and Seebeck[16] coefficients at low temperature have demonstrated that the Fermi surface (FS) of underdoped YBCO undergoes a reconstruction at low temperature and consists of at least one electron pocket. A comparative study of thermoelectric transport in underdoped YBCO and in La$_{1.8-x}$Eu$_{0.2}$Sr$_x$CuO$_4$ (a cuprate where stripe order is well established) argues in favor of a charge stripe order causing reconstruction of the FS at low temperature[17]. High field nuclear magnetic resonance (NMR) measurements have indeed revealed that the translational symmetry of the CuO$_2$ planes in YBCO is broken by the emergence of a unidirectional modulation of the charge density at low temperature, above a threshold magnetic field[4]. In zero field, long range charge fluctuations in YBCO were recently observed with resonant soft x-ray scattering (RSXS) up to 150 K and 160 K for $p = 0.11$ (ref. 5) and $p = 0.133$ (ref. 6) respectively, while hard x-ray scattering experiments suggest that a charge order develops below 135 K for $p = 0.12$ (ref. 7). All measurements identify charge fluctuations at two wavevectors corresponding to an incommensurate period $\approx 3.2$ lattice units. The identification of a thermodynamic phase transition is thus important to determine where long range charge order exists in the phase diagram and particularly whether static order only occurs in high magnetic fields.

Here we report sound velocity measurements, a thermodynamic probe, in magnetic fields large enough to suppress superconductivity. The sound velocity is defined as $v_s = \sqrt{\dfrac{c_{ij}}{\rho}}$, where $\rho$ is the density of the material, $c_{ij} = \dfrac{\partial^2 F}{\partial \varepsilon_i \partial \varepsilon_j}$ (ref. 18), $F$ the free energy and $\varepsilon_i$ the strain along direction i. Changes in the elastic constants $c_{ij}$ are expected whenever a strain dependent phase transition occurs. Owing to their high sensitivity, sound velocity measurements are a powerful probe for detecting such phase transitions, in particular charge ordering in strongly correlated electron system[19].

We have measured several elastic constants (see Supplementary table S1 for the description of the elastic modes) in high magnetic fields in an underdoped YBCO$_{6.55}$ sample with $T_c = 60.7$ K corresponding to a hole doping $p = 0.108$[20]. Fig. 1a and 1b show the field dependence of the relative variation of the sound velocity $\Delta v_s/v_s$ corresponding to the c$_{11}$ mode, at different temperatures. At $T$=4.2 K, the softening of the elastic constant at $B_m \approx 20$ T corresponds to the first order melting transition from a vortex lattice to a vortex liquid[21,22] (see Supplementary for more details). At $T$=29.5 K, this anomaly shifts to lower field ($B_m \approx 5$ T) and since the pinning potential becomes less effective, the magnitude of the change of $c_{11}$ at the melting transition becomes smaller[23]. At $T$ = 29.5 K and above $B_m$, a sudden increase of the elastic constant can clearly be resolved at $B_{co}$ = 18 T which corresponds to a thermodynamic signature of a phase transition. While $B_{co}$ is almost temperature independent at low temperature, it increases rapidly between 35 and 50 K (see red arrows in Fig. 1b). Above $T$ = 50 K, no change of $c_{11}$ can be resolved up to the highest field. Because of the difference in the temperature dependence of $B_m$ and $B_{co}$, the phase transition at $B_{co}$ can not originate from vortices. Fig. 2 shows the phase diagram where $B_m$ and $B_{co}$ are plotted as a function of temperature. The identification of this phase stabilized by the magnetic field above $B_{co}$ is straightforward. High field NMR measurements in YBCO at similar doping have shown that charge order develops above a threshold field $B_{co}$ >15 T and below $T_{co}^{RMN}$ = 50 ±10 K (ref. 4). Given the similar field and temperature scales, it is natural to attribute the anomaly seen in the elastic constant at $B_{co}$ to the thermodynamic transition towards the static charge order.

The phase diagram of Fig. 2 shares common features with the theoretical phase diagram of superconductivity in competition with a density wave order[24] (see discussion in the Supplementary). For $T$ below 40 K or so, static charge order sets in only above a threshold field of 18 T, akin to the situation in La$_{2-x}$Sr$_x$CuO$_4$ (x = 0.145) where a magnetic field is necessary to destabilize superconductivity and to drive the system to a magnetically ordered state[11]. Close to the onset temperature of static charge order, $T_{co}$, the $B_{co}$ line connects with the temperature $T_0^{R_H}$ where the Hall coefficient $R_H$(T) changes sign, which coincides with the onset of charge order seen in NMR[4]. $R_H$(T) of YBCO ($p$ = 0.108) changes sign at a characteristic temperature $T_0^{R_H}$ = 45 K, which is field independent above 30 T [15,25] (see blue symbols in Fig. 2). This indicates that, in this part of the phase diagram, superconducting fluctuations have no significant effect on charge order.

We now turn to the analysis of the symmetry of the charge modulation. In the framework of the Landau theory of phase transitions, an anomaly in the elastic constant occurs at a phase transition only if a coupling in the free energy $F_c = g_{mn} Q^m \varepsilon^n$ (where $m$ and $n$ are integers, $g_{mn}$ is a coupling constant) between the order parameter $Q$ and the strain $\varepsilon$ is symmetry allowed, i.e. only if $Q^m$ and $\varepsilon^n$ transform according to the same irreducible representation[26]. In Fig. 3 we compare the field dependence at $T = 20$ K of four different modes $c_{11}$, $c_{44}$, $c_{55}$ and $c_{66}$ which display an anomaly at $B_{co}$. To explain the presence of such coupling for all these modes, we rely on group theory arguments. YBCO is an orthorhombic system (point group $D_{2h}$), and given the even character of the strains we only have to consider the character table of point group $D_2$ shown in Table 1.

To represent the different symmetric charge modulations that transform according to each irreducible representation of the point group $D_2$ and to determine to which acoustic mode they couple, we will follow the procedure of Goto and Lüthi (ref. 19) and use the projection operator in the basis defined by the four Cu atoms at the corner of the CuO$_2$ plane (see S. I.). This simplification is justified since the charge order is an intrinsic property of the CuO$_2$ plane[4,6]. The four kinds of symmetric charge distribution allowed in this basis are sketched in Fig. 4. A unidirectional charge modulation along the $a$-axis should induce an anomaly in $c_{44}$ but not in $c_{55}$. The very similar field dependence of $c_{44}$ and $c_{55}$ around $B_{co}$ (see Fig. 3) indicates a two-dimensional charge modulation both along the $a$-axis ($Q_a$) and along the $b$-axis ($Q_b$). Since $B_{1g} = B_{2g} \otimes B_{3g}$ and $\varepsilon_6$ transforms according to $B_1$, an additional coupling $\beta_{ab} Q_a Q_b \varepsilon_6$ explains the anomaly seen in $c_{66}$. A similar coupling exists for $c_{11}$ since $A_{1g} = B_{ig} \otimes B_{ig}$ (i=1, 2 or 3).

As a thermodynamic bulk probe with a typical energy scale below 1 µeV, sound velocity measurements establish that static charge order sets in below $T_{co} \approx 45$ K in magnetic fields above $B_{co} = 18$ T in YBCO at a doping level $p = 0.108$. For a nearby doping $p = 0.11$ (non ortho II ordered sample), zero field RSXS measurements show that charge fluctuations appear at $T = 150$ K[5]. The fluctuating character of the charge order at high temperature is consistent with recent resonant ultrasonic measurements in YBCO at similar doping level. While anomalies in the sound velocity are detected at the pseudogap temperature, $T^* \approx 240$ K, and at the superconducting transition temperature $T_c$, no additional phase transition is observed in zero field[27]. An important aspect of theses fluctuations is their two-dimensional character inferred from both RSXS[5,6] and x-ray[7] measurements. However those techniques cannot determine whether there is a single charge density wave (CDW) with biaxial modulation or if there are domains of two uniaxial CDW. From sound velocity measurements, the latter scenario seems unlikely since for the anomaly in $c_{66}$ to exist, charge modulations along the two directions should in principle coexist within the same sample volume. The similar dimensionality observed in our study and in x-ray measurements supports the scenario where the static charge ordering occurring at low temperature and finite magnetic field is a consequence of the freezing of the charge fluctuations observed in zero field at high temperature. This dimensionality seems to differ from NMR measurements, which have been interpreted as a unidirectional modulation of the charge density along the $a$-axis[4]. However an additional charge modulation along the $b$-axis, as reported here, is not excluded from the NMR spectra. The dimensionality of the charge order has a strong impact on the topology of the reconstructed Fermi surface. In the absence of spin order, as inferred from NMR[4], unidirectionnal order by itself cannot produce electron pockets in the reconstructed Fermi surface[28], unless one assumes that the Fermi surface in the high temperature already breaks $C_4$ symmetry[29]. On the other hand, no particular assumption is needed to produce electron pockets when the Fermi surface is folded along two perpendicular directions[9].

1. Kivelson, S. A. *et al*. How to detect fluctuating stripes in the high-temperature superconductors. *Rev. Mod. Phys.* 75, 1201 (2003).

2. Vojta, M. Lattice symmetry breaking in cuprate superconductors: Stripes, nematics, and superconductivity. *Adv. Phys.* **58**, 699 (2009).

3. Taillefer, L. Scattering and Pairing in Cuprate Superconductors. *Annu. Rev. Condens.Matter Phys.* **1**, 51-70 (2010).

4. Wu, T. *et al*. Magnetic-field-induced charge-stripe order in the high-temperature superconductor $YBa_2Cu_3O_y$. *Nature* **477**, 191-194 (2011).

5. Ghiringhelli, G. *et al*. Long-range incommensurate charge fluctuations in $(Y,Nd)Ba_2Cu_3O_{6+x}$. *Science* **337**, 821 (2012).

6. Achkar, A. J. *et al.* Distinct charge orders in the planes and chains of ortho-III ordered $YBa_2Cu_3O_{6+\delta}$ identified by resonant x-ray scattering. Preprint at <http://arXiv.org/abs/ 1207.3667>, (2012).

7. Chang, J. *et al*. Direct observation of competition between superconductivity and charge density wave order in $YBa_2Cu_3O_y$. Preprint at <http://arXiv.org/abs/ 1206.4333>, (2012).

8. Yao, H. *et al*. Fermi-surface reconstruction in a smectic phase of a high-temperature superconductor. *Phys. Rev. B* **84**, 012507 (2011).

9. Sebastian, S. *et al*. Quantum oscillations from nodal bilayer magnetic breakdown in the underdoped high temperature superconductor $YBa_2Cu_3O_{6+x}$. *Phys. Rev. Lett.* **108**, 196403 (2012).

10. Lake, B. *et al*. Antiferromagnetic order induced by an applied magnetic field in a high-temperature superconductor. *Nature* **415**, 299-302 (2002).

11. Chang, J. *et al.* Tuning competing orders in $La_{2-x}Sr_xCuO_4$ cuprate superconductors by the application of an external magnetic field. *Phys. Rev. B* **78**, 104525 (2008).

12. Wen, J. *et al*. Uniaxial linear resistivity of superconducting $La_{1.905}Ba_{0.095}CuO_4$ induced by an external magnetic field. *Phys. Rev. B* **85**, 134513 (2012).

13. Doiron-Leyraud, N. *et al.* Quantum critical point for stripe order : an organizing principle of cuprate superconductivity. *Physica C* **481**, 161 (2012).

14. Doiron-Leyraud, N *et al.* Quantum oscillations and the Fermi surface in an underdoped high-$T_c$ superconductor. *Nature* **447**, 565-568 (2007).

15. LeBoeuf, D. *et al.* Electron pockets in the Fermi surface of hole-doped high-$T_c$ superconductors. *Nature* **450**, 533-536 (2007).

16. Chang, J. *et al.* Nernst and Seebeck coefficients of the cuprate superconductor $YBa_2Cu_3O_{6.67}$ : a study of Fermi surface reconstruction. *Phys. Rev. Lett.* **104**, 057005 (2010).

17. Laliberté, F. *et al.* Fermi-surface reconstruction by stripe order in cuprate superconductors. *Nature Comm.* **2**, 432 (2011).

18. Lüthi, B. Physical acoustics in the solid state. Springer series in the solid state science 148 (2005)

19. Goto, T and Lüthi, B. Charge ordering, charge fluctuations and lattice effects in strongly correlated electron systems. *Advances in Physics* **52**, 67-118 (2003).

20. Liang, R. *et al*. Evaluation of $CuO_2$ plane hole doping in $YBa_2Cu_3O_{6+x}$ single crystals. *Phys. Rev. B* **73**, 180505 (2006).

21. Liang, R. *et al*. Discontinuity of reversible magnetization in untwinned YBCO single crystals at the first order vortex melting transition. *Phys. Rev. Lett*. **76,**:835 (1996)

22. Ramshaw, B. *et al*. Vortex lattice melting and $H_{c2}$ in underdoped $YBa_2Cu_3O_y$. Preprint at <http://arXiv.org/abs/ 1209.1655>, (2012).

23. Pankert, J. *et al*. Ultrasonic attenuation by the vortex lattice of high-$T_c$ superconductors. *Phys. Rev. Lett*. **65**, 3052 (1990).

24. Moon, E. G. and Sachdev, S. Competition between spin density wave order and superconductivity in the underdoped cuprates. *Phys. Rev. B* **80**, 035117 (2009).

25. LeBoeuf, D. *et al*. Lifshitz critical point in the cuprate superconductor $YBa_2Cu_3O_y$ from high-field Hall effect measurements. *Phys. Rev. B* **83**, 054506 (2011).

26. Rehwald, W. The study of structural phase transitions by means of ultrasonic experiments. *Adv. Phys*. **22**, 721 (1973)

27. Shekhter, A. Ultrasonic signatures at the superconducting and the pseudogap phase boundaries in YBCO cuprates. Preprint at <http://arXiv.org/abs/ 1208.5810>, (2012).

28. A. J. Millis and M. R. Norman. Antiphase stripe order as the origin of electron pockets observed in 1/8-hole-doped cuprates. *Phys. Rev. B* **76**, 220503 (2007)

29. H. Yao *et al.* Fermi-surface reconstruction in a smectic phase of a high-temperature superconductor. *Phys. Rev. B* **84**, 012507 (2011)

**Acknowledgments** We thank M.-H. Julien, S. Kivelson, B. Lüthi, R. Ramazashvili, G. Rikken, L. Taillefer, B. Vignolle, M. Vojta and S. Zherlitsyn for useful discussions. We acknowledge experimental supports from A. Mari, D. Rickel and the LNCMI staff. Research support was provided by the French ANR SUPERFIELD, Euromagnet II, the Canadian Institute for Advanced Research and the Natural Science and Engineering Research Council.

Correspondence and requests for materials should be addressed to D. L. (david.leboeuf@lncmi.cnrs.fr) and C.P. (cyril.proust@lncmi.cnrs.fr).

**Table 1 | Character table of point group *D*₂**

Character table, basis functions and transformation properties of the strains for the point group *D*₂ appropriate for YBCO. The z, y and x-axis correspond to the crystalline c, b and a-axis of the YBCO structure respectively.

| Irr. Rep. | $E$ | $C_2^z$ | $C_2^y$ | $C_2^x$ | Basis functions | Symmetric strains |
|-----------|-----|---------|---------|---------|-----------------|-------------------|
| $A_{1g}$ | 1 | 1 | 1 | 1 | $x^2$ , $y^2$ , $z^2$ | Volume strains : $\varepsilon_1$ |
| $B_{1g}$ | 1 | 1 | -1 | -1 | z, xy | $\varepsilon_6$ |
| $B_{2g}$ | 1 | -1 | 1 | -1 | y, xz | $\varepsilon_5$ |
| $B_{3g}$ | 1 | -1 | -1 | 1 | x, yz | $\varepsilon_4$ |

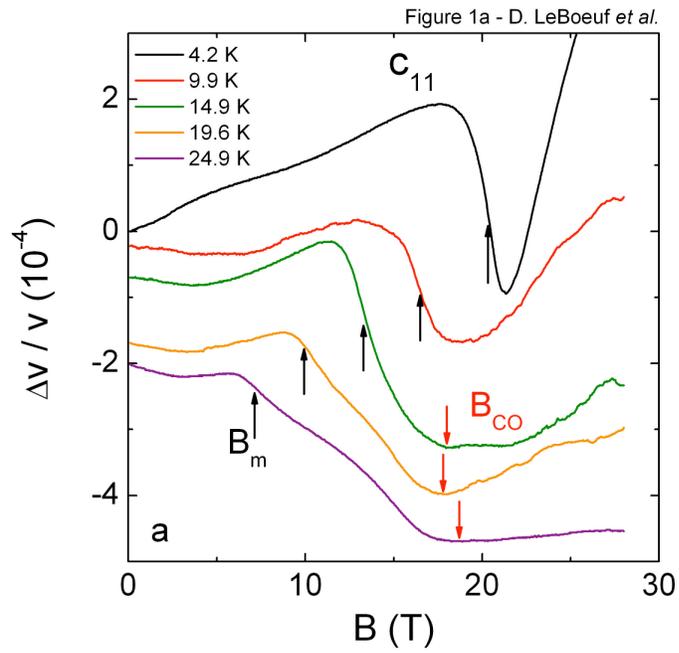

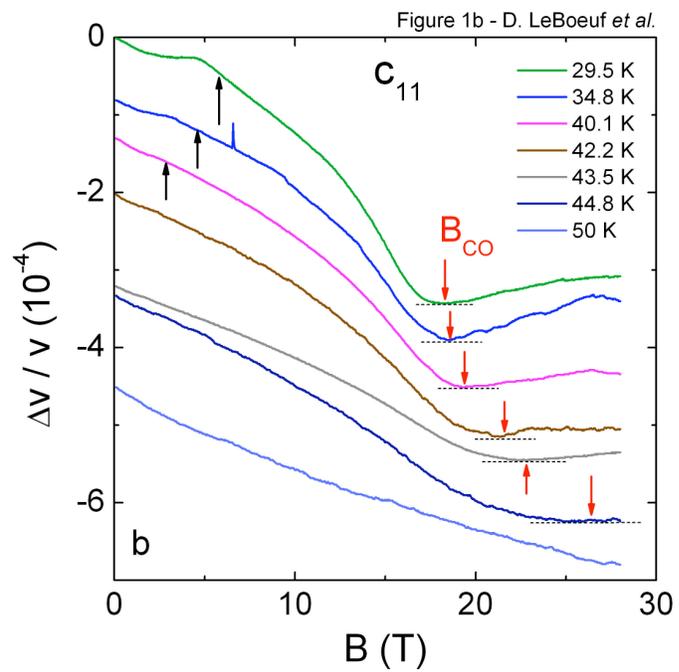

**Figure 1 | Field dependence of the sound velocity in underdoped YBa$_2$Cu$_3$O$_y$.**

Field dependence of the longitudinal mode $c_{11}$ (propagation $q$ and polarization $u$ of the sound wave along $a$-axis) at different temperatures a) from $T$ = 4 K to $T$ = 24.9 K, and b) from $T$ = 29.5 K to $T$ = 44.8 K. Curves are shifted for clarity. The measurements have been performed in static magnetic field up to 28 T. Black arrows indicate the field $B_m$ corresponding to the vortex lattice melting. At low temperature, the loss of the vortex lattice rigidity contribution to the sound velocity can be estimated and is in agreement with previous studies (see Supplementary). For $T$ > 40 K, $B_m$ cannot be resolved. Red arrows indicate the field $B_{co}$ where the charge order phase transition occurs. This transition is not related to vortex physics since it is also seen in acoustic modes $c_{44}$ and $c_{55}$ (see Fig. 3 and Fig. S3) which are insensitive to the flux line lattice since those modes involve atomic motions parallel to the vortex flux lines (u // H // c).

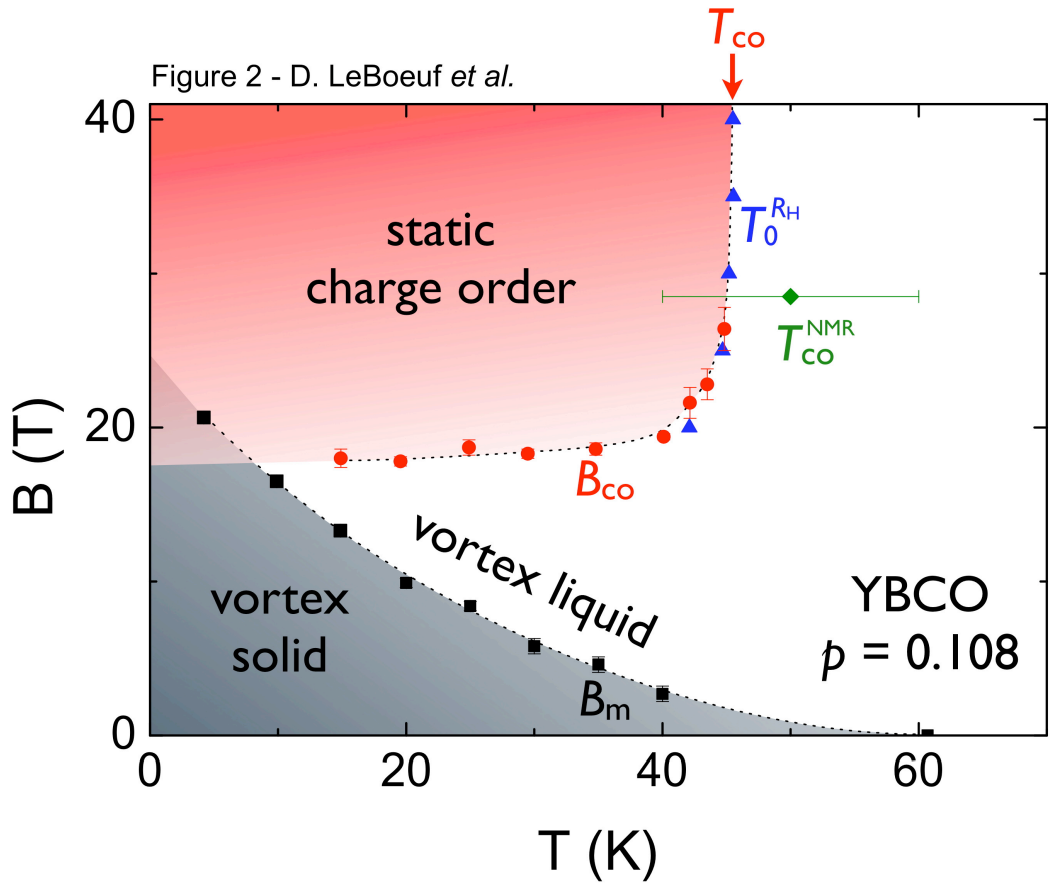

**Figure 2 | Thermodynamic phase diagram.**

Magnetic field - temperature phase diagram of underdoped YBCO ($p = 0.108$) obtained from the anomalies seen in the elastic constant $c_{11}$ (see Fig. 1). Black squares indicate the transition from a vortex lattice to a vortex liquid at $B_m$, which cannot be resolved above 40 K. Red circles correspond to the phase transition towards static charge order at $B_{co}$, as observed in $c_{11}$. The charge order transition is almost temperature independent up to ≈ 40 K. Above 40 K the transition rises steeply and extrapolate to a vertical behaviour at higher fields and connect to $T_0^{R_H}$ (blue symbols), the sign change temperature of the Hall coefficient. In the Supplementary, we argue that the overall behavior of the charge order phase boundary in this *B-T* diagram is consistent with a theoretical model of superconductivity in competition with a density wave state[24]. The green diamond is the temperature $T_{co}^{NMR}$ = 50 ± 10 K where NMR experiments detect the onset of a charge modulation at a field $B$ = 28.5 T in YBCO at doping $p$ = 0.11[4]. Within the error bars, this onset temperature agrees with our findings. Dashed lines are guides to the eye.

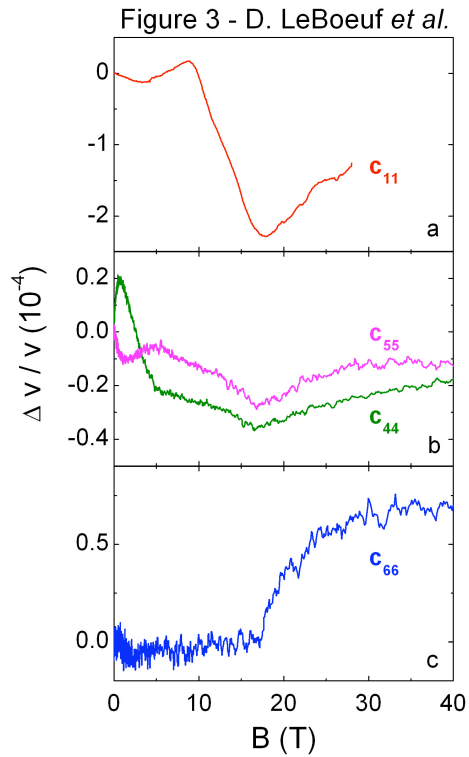

**Figure 3 | Charge order transition seen in different modes.**

Field dependence of different elastic constants of YBCO (p = 0.108) at $T$ = 20 K a) longitudinal mode $c_{11}$, b) $c_{44}$ ($q//b$, $u//c$) in green and $c_{55}$ ($q//a$, $u//c$) in magenta, and c) $c_{66}$ ($q//b$, $u//a$). $c_{44}$ and $c_{55}$ have similar amplitude and field dependence at the transition $B_{co}$ towards the charge order indicating the two-dimensional character of the charge density modulation. The anomalies seen in $c_{66}$ and $c_{11}$ are larger than in $c_{44}$ and $c_{55}$. This is probably due to a greater dependence of charge order with respect to in-plane strains than out of plane strains.

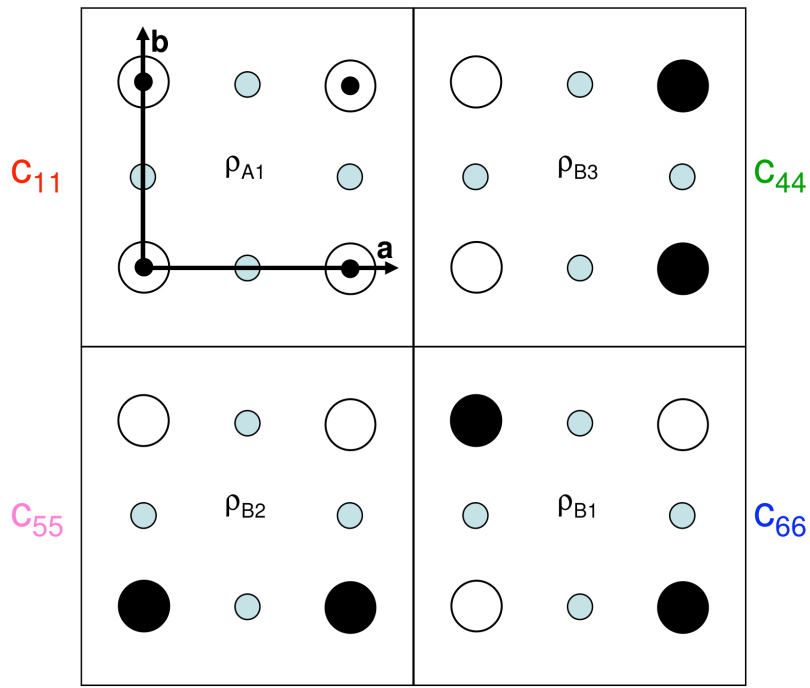

**Figure 4 |. Charge distribution in the CuO₂ plane**

Sketches of the different charge distributions in the CuO₂ plane associated with the four irreversible representations of the point group $D_2$ (see Table 1). Black (blue) circles represent Cu (O) atoms. Filled (empty) black circles correspond to higher (lower) charge density. The periodicity of the modulation is arbitrary. The charge distribution is obtained by applying the projection operator in the CuO₂ plane of the orthorhombic structure of YBCO (see Supplementary). A unidirectional charge order with a modulation along the *a(b-)*-axis transforms according to $B_{3g}$ ($B_{2g}$) and couples linearly with $c_{44}$ ($c_{55}$). Since both $c_{44}$ and $c_{55}$ show an anomaly at $B_{co}$, we conclude that the charge modulation is bidirectional.

# Thermodynamic phase diagram of static charge order in underdoped YBa$_2$Cu$_3$O$_y$

David LeBoeuf, S. Krämer, W.N. Hardy, Ruixing Liang, D.A. Bonn, and Cyril Proust

## EXPERIMENTAL METHODS

The sample is an uncut, detwinned crystal of YBa$_2$Cu$_3$O$_y$ grown in a nonreactive BaZrO$_3$ crucible from high-purity starting materials[30] with oxygen content y = 6.55. The oxygen atoms in the CuO chains were made to order into the ortho-II superstructure. The hole concentration (doping) $p$ = 0.108 of the samples was determined from the superconducting transition temperature $T_c$ = 60.7 K (ref. 20) defined as the midpoint of the magnetization transition.

Sample dimensions are (2.53 × 1.25 × 0.2) mm$^3$ [(length) × (width) × (thickness)].

Ultrasonic waves in the range 150 - 350 MHz were generated using commercial LiNbO$_3$ transducers glued to the surface of the sample. 36° Y-cut transducers were used to generate longitudinal waves and 41° X-cut to generate shear waves. A standard pulse-echo technique with phase comparison was used, either in phase-locked mode in static magnetic field or in frequency-locked mode in pulsed magnetic field.

The magnetic field was applied along the $c$-axis of the orthorhombic structure, perpendicular to the CuO$_2$ planes.

The longitudinal mode $c_{11}$ was measured in a static 20 MW resistive magnet at the LNCMI in Grenoble providing up to 28 T. All the other modes were measured in pulsed magnetic field up to 60 T in a resistive magnet at the LNCMI in Toulouse. In static magnetic field, the slow field sweep rate allows time-averaging of the signal, whereas it is not possible in pulsed magnetic field. Consequently, the sensitivity in pulsed magnetic field is lower than in static magnetic field (see Fig. 3 and Fig. S3).

**Table S1 | Propagation direction and polarization of the different acoustic modes**

| Mode | Propagation vector $q$ | Polarisation vector $u$ |
|------|------------------------|--------------------------|
| $c_{11}$ | [1,0,0] | [1,0,0] |
| $c_{44}$ | [0,1,0] | [0,0,1] |
| $c_{55}$ | [1,0,0] | [0,0,1] |
| $c_{66}$ | [0,1,0] | [1,0,0] |

## IDENTIFICATION OF $B_m$ AND $B_{co}$

To find the right criteria to locate $B_m$ in the sound velocity measurements, we have measured in the same experimental conditions the field dependence of the sound velocity and of the resistance in another YBCO sample with similar doping for which gold pads were evaporated on the surface. As seen in Fig. S1c, the appearance of a finite resistance at $B_m$ corresponds to the inflection point in $\Delta v / v$ (see Fig. S1a and the derivative in Fig. S1b).

Fig. S2 shows the derivative of $\Delta v / v$ with respect to $B$ at $T = 29.5$ K for the longitudinal mode $c_{11}$. This derivative features two inflection points : the first one is related to the melting of the vortex lattice as shown in Fig. S1, and the second inflection at higher field precedes the charge order transition at $B_{co}$, that is identified as the minimum in $c_{11}$. Fig. S2 clearly demonstrates the presence of two phase transitions in the field dependence of the longitudinal mode $c_{11}$ at $T = 29.5$ K.

**Figure S1 | Identification of the vortex lattice melting transition $B_m$**

Data taken in another YBCO sample with similar doping. **a)** Sound velocity in the $c_{11}$ mode at $T = 20$ K as a function of magnetic field. For $B < B_m$ the signal has a hysteretic behavior as expected in the vortex solid phase. **b)** Derivative of $\Delta v / v$ with respect to $B$. The inflection point of $\Delta v / v$ defines $B_m$. **c)** Field dependence of the interlayer resistivity $R_c$ measured in the same sample and in the same experimental conditions.

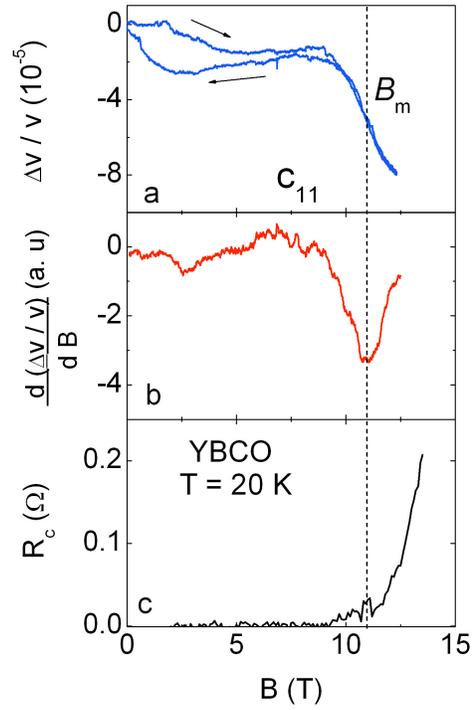

**Figure S2 | Distinction between $B_m$ and $B_{co}$**

**a)** $\Delta v / v$ for the $c_{11}$ mode as a function of magnetic field for $T = 29.5$ K. **b)** Derivative of $\Delta v / v$ with respect to magnetic field for $T = 29.5$ K.

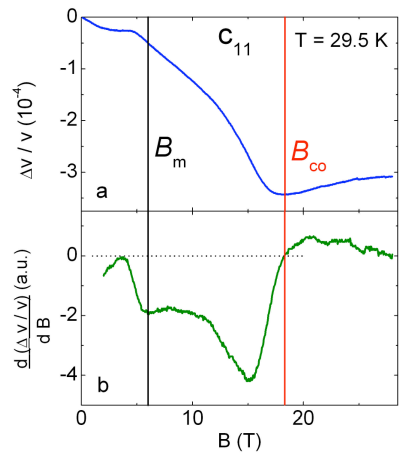

## TRANSVERSE MODES : $c_{44}$, $c_{55}$ and $c_{66}$

Fig. S3 shows the field dependence of the three transverse modes $c_{44}$, $c_{55}$ and $c_{66}$ that transform according the representation $B_3$, $B_2$ and $B_1$ respectively. All these modes show an anomaly at the charge order transition $B_{co}$. Fig. S4 shows the temperature dependence of $B_{co}$ deduced from the three transverse modes. It is in agreement with the charge order transition observed in the longitudinal mode $c_{11}$. The irreversibility line $B_m$ does not appear in $c_{55}$ and $c_{44}$, and is not resolved in $c_{66}$ as discussed in the next section.

**Figure S3 | Field dependence of transverse modes $c_{44}$, $c_{55}$ and $c_{66}$**

Field dependence of the transverse mode **a)** $c_{44}$ **b)** $c_{55}$ and **c)** $c_{66}$ at different temperatures. Curves are shifted for clarity.

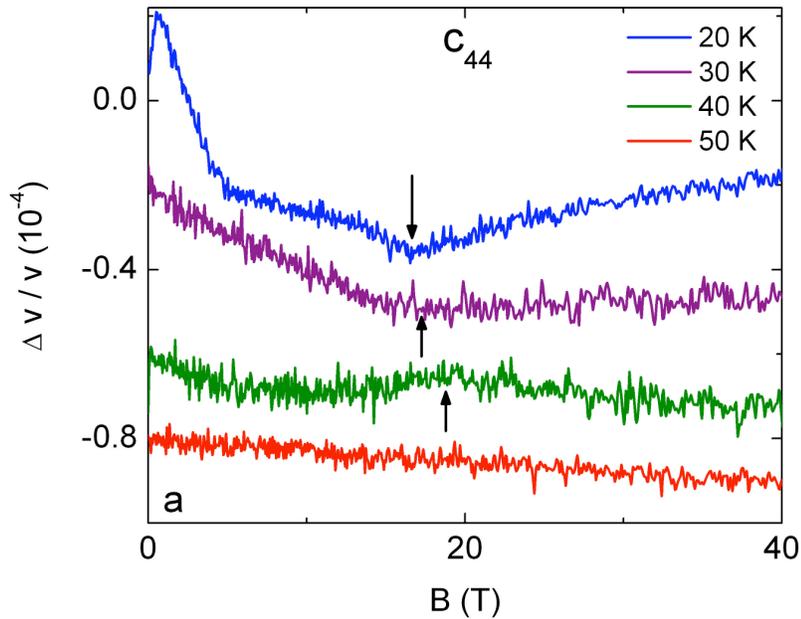

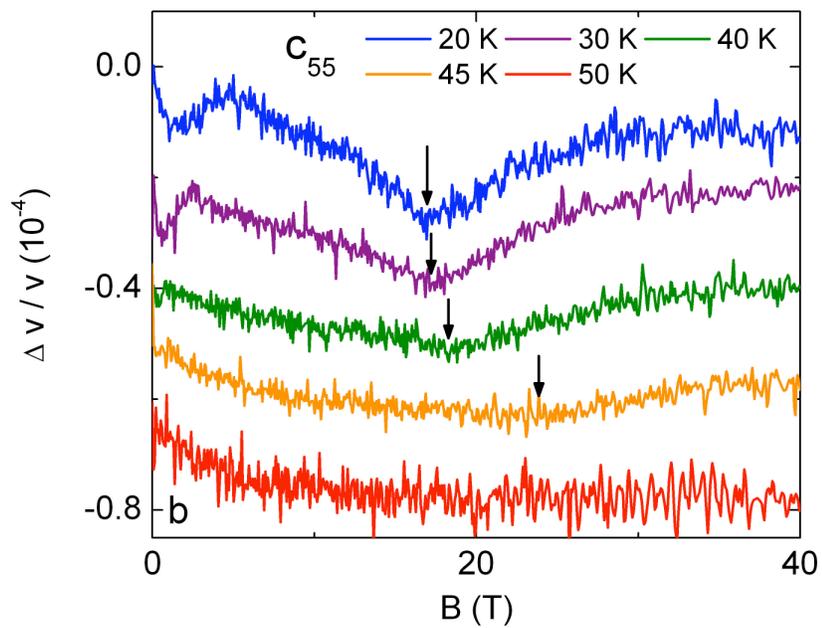

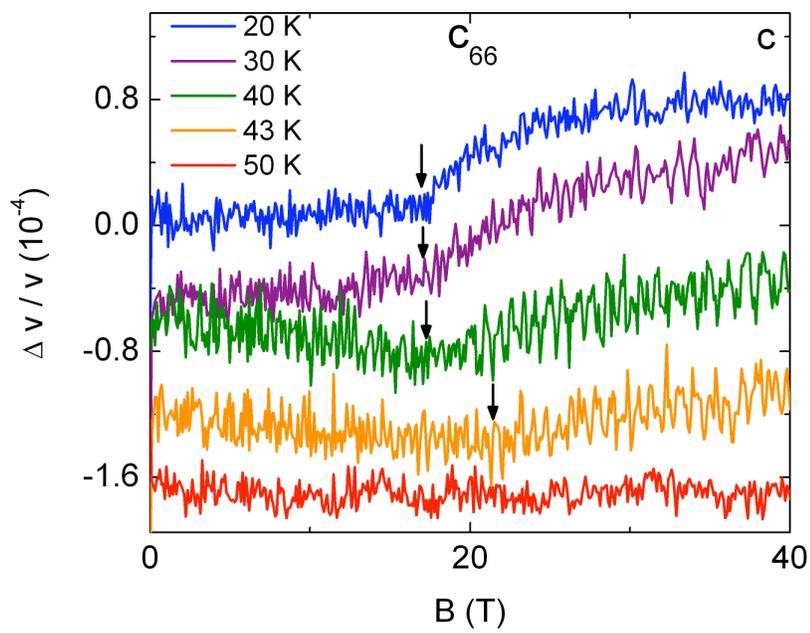

**Figure S4 | Phase diagram deduced from c₄₄, c₅₅, c₆₆, and c₁₁**

Comparison of the onset field of charge order $B_{co}$ deduced from the anomaly seen in different modes. Red circles : $c_{11}$. Blue triangles : $c_{66}$. Green diamonds : $c_{55}$. Magenta triangles : $c_{44}$. Black squares show the vortex transition $B_m$ observed in $c_{11}$. Dashed lines are guide to the eye.

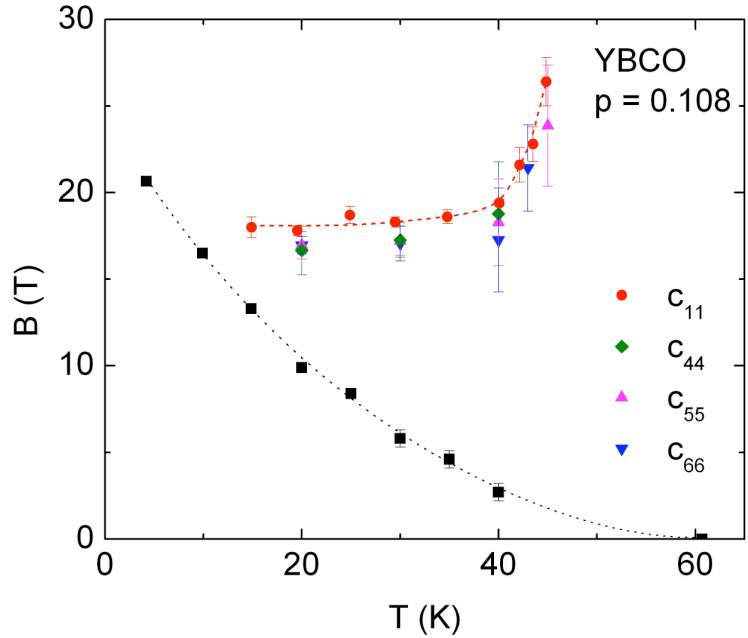

## VORTEX LATTICE CONTRIBUTION TO THE SOUND VELOCITY

The vortex lattice is characterized by three independent elastic moduli, namely, $c_{11}^v$ the compression modulus, $c_{44}^v$ the tilt modulus and $c_{66}^v$ the shear modulus[31]. In the presence of vortex pinning, the vortex lattice moduli $c_{ii}^v$ add to the crystal lattice moduli $c_{ij}$ in the measured sound velocity[32]: $\rho V_S^2 = c_{ii}^v + c_{ij}$

The vortex lattice modulus contribution $c_{ii}^v$ to $V_S$ is zero for $u /\!/ B$, as in $c_{44}$ and $c_{55}$ (see Table S1). However, for $c_{11}$ and $c_{66}$ where $u \perp B$ (see Table S1), the vortex lattice contribution is :

$$\left| c_{11}^v \right| \approx \frac{B^2}{\mu_0} \text{ and } \left| c_{66}^v \right| \approx \frac{\phi_0 B}{16 \pi \mu_0 \lambda^2}$$ where $\lambda$ is the magnetic penetration depth, $\mu_0$ the magnetic

permeability of free space and $\phi_0$ the magnetic flux quantum[31].

When the vortex lattice melts at $B_m$, $c_{ii}^v$ is strongly reduced. Consequently one sees a drop in $\Delta V_S / V_S$ at $B_m$ (see Fig. 1 a, and Fig. S1). At $T = 4.2$ K, $B_m \approx 20$ T which gives $c_{11}^v \approx 318$ MPa. With $c_{11} = 231$ GPa measured at room temperature[33], we estimate $\left| \Delta V_s / V_s \right| = c_{11}^v / 2c_{11} \approx 7 \times 10^{-4}$. Note that this estimation is an upper limit, since $c_{11}$ increases by a few percents upon cooling. This estimation is in good rough agreement with our measurements of the longitudinal mode $c_{11}$ at $T = 4.2$ K, which gives $\left| \Delta V_s / V_s \right| = 3 \times 10^{-4}$ (see Fig. 1 a). Earlier reports already noticed that the experimental values for $c_{ii}^v$ can be reduced by as much as 70% in comparison to the theoretical expectation[34]. The drop in $\Delta V_S / V_S$ at $B_m$ observed in $c_{11}$ becomes gradually smaller as the temperature increases (see Fig. 1 a). This temperature dependence of $c_{ii}^v$ can be explained with the theory of thermally assisted flux flow[23].

For the vortex lattice shear modulus, we get $c_{66}^v \approx 22$ kPa for $B_m \approx 20$ T and using $\lambda \approx 170$ nm (ref. 35). With $c_{66} = 95$ GPa at room temperature[33], we estimate $\left| \Delta V_s / V_s \right| = c_{66}^v / 2c_{66} \approx 10^{-7}$, well below our experimental resolution in pulsed fields.

# CALCULATION OF THE SYMMETRIC CHARGE DISTRIBUTIONS

In order to determine the symmetric charge distributions $\rho_\Gamma$ that transform according to each irreducible representation $\Gamma$ of the point group $D_2$ we use the projection operator :

$$P_\Gamma = \frac{1}{d} \sum_R \chi^\Gamma(R) P_R$$

where $R$ is a symmetry of the group, $\chi^\Gamma(R)$ is the character of the symmetry $R$ in the representation $\Gamma$ and $P_R$ the operator of the symmetry $R$[19]. Note that this definition of $P_\Gamma$ is only valid for 1D representations. We apply $P_\Gamma$ to the basis defined by the charge distribution $\rho_i$ at the four Cu atom sites (1-4), as shown in Fig. S5. This basis is orthorhombic in the case of YBCO, and given the even character of the strain, we have to consider the character table of point group $D_2$. If we apply for instance the projection operator in the representation $A_1$ on the charge distribution on site 1, namely $\rho_1$, we have:

$$\rho_{A_1} = P_{A_1} \rho_1 = \frac{1}{4} \left\{ P_E \rho_1 + C_2^z \rho_1 + C_2^y \rho_1 + C_2^x \rho_1 \right\} = \frac{1}{4} \left\{ \rho_1 + \rho_3 + \rho_2 + \rho_4 \right\}$$

This result shows that the symmetric charge distribution that transforms according to $A_1$ corresponds to a uniform and equal distribution $\rho_i$ on the four sites (1-4), as shown in Fig. 4, upper left pannel. This result is expected since the representation $A_1$ conserves the symmetry of the system. We use the projection operator in the same way in order to determine $\rho_\Gamma$ for the other irreducible representations of $D_2$ :

$$\rho_{B_1} = \rho_1 - \rho_2 + \rho_3 - \rho_4$$
$$\rho_{B_2} = \rho_1 + \rho_2 - \rho_3 - \rho_4$$
$$\rho_{B_3} = \rho_1 - \rho_2 - \rho_3 + \rho_4$$

From this calculation, the four kinds of charge distribution $\rho_\Gamma$ that transform according to each representation $\Gamma$ of the point group $D_2$ can be deduced. A schematic representation of the four $\rho_\Gamma$ is given in Fig. 4. The calculation is performed by assuming that the charge is centered on the Cu sites (see Fig. S5). This assumption does not influence the conclusions drawn in this paper. This calculation can also be performed taking into account the oxygen atoms and also give that $c_{44}$ couples linearly with a charge modulation along the a-axis and $c_{55}$ couples linearly with a charge modulation along the b-axis.

**Figure S5 | CuO₂ plane of the YBCO orthorhombic structure.**

Black circle : Cu atoms. Blue circles : O atoms. The Cu atoms are labeled 1-4. These labels are used for the symmetry analysis described above.

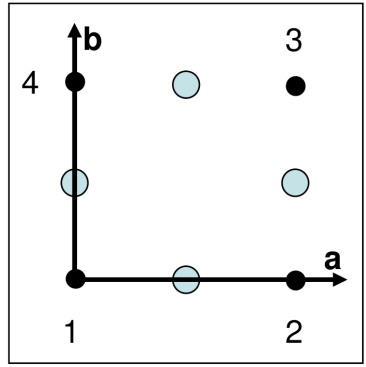

# THERMODYNAMIC PHASE TRANSITION AND ELASTIC CONSTANTS

The coupling between the order parameter $Q$ and the strain ε has been introduced as an additional term $F_c = g_{mn} Q^m \varepsilon^n$ in the free energy $F$ of the system. Here we shall give a perhaps more intuitive way of introducing this coupling, in order to discuss what kind of anomalies in the sound velocity are expected at a phase transition. For a second order phase transition, the Ginzburg-Landau free energy is:

$F = a(T)Q^2 + \dfrac{1}{2}bQ^4 = -\Phi_0(T-T_c)^2$ , with $\Phi_0$ a condensation energy and $T_c$ the critical temperature for the phase transition. If one assumes that both $\Phi_0$ and $T_c$ are strain-dependent, one can then calculate the change in elastic constant at the transition:

$\Delta c_{ii} = \dfrac{\partial^2 F}{\partial \varepsilon_i^2} = -2\Phi_0\left(\dfrac{\partial T_c}{\partial \varepsilon_i}\right)^2 - \left[2\Phi_0\dfrac{\partial^2 T_c}{\partial \varepsilon_i^2} + 4\dfrac{\partial \Phi_0}{\partial \varepsilon_i}\dfrac{\partial T_c}{\partial \varepsilon_i}\right](T_c-T) - \dfrac{\partial^2 \Phi_0}{\partial \varepsilon_i^2}(T_c-T)^2$

The first term results in a step in the sound velocity at the transition. The second term gives rise to a change of slope and the third term to a change in the curvature of the sound velocity. Depending on the symmetry of the system, the nature of the transition and on the acoustic mode used to probe that transition, different kind of anomalies in the sound velocity can be expected[18]. At the vortex lattice melting transition $B_m$, the loss of the vortex lattice contribution should in theory result in a step-like anomaly in the sound velocity. The transition observed here in more rounded probably due to the effect of disorder. The charge order transition $B_{co}$ has a significantly different nature and consequently gives rise to different kind of anomalies in the sound velocity.

## PHASE DIAGRAM OF COMPETING ORDER

Fig. S6 a) reproduces the phase diagram of Fig. 2. It has been shown empirically that $T_0 = T_{co}$, $T_{co}$ being the onset of charge order seen in NMR[4]. This is also in agreement with our data the onset of static charge order seen in the sound velocity overlaps with and extrapolates remarkably well to $T_0$.

We compare the experimental phase diagram (Fig. S6 a) with the $T = 0$, theoretical phase diagram for a system in which superconductivity competes with another order[24] (Fig. S6 b). The parameter $t$ controls the strength of the competing order and acts similarly as the doping $p$[24] (decreasing $t$ favors the competing order). This theoretical diagram predicts two characteristic features that are in agreement with our experimental diagram. First, for a doping such that $t_c < t < t_c(0)$, as in the case of LSCO $p = 0.145$[11], application of an external threshold field $B_{co}$ is necessary to observe static charge order. This is in agreement with our findings, since the transition towards static order is seen at a finite field. Remarkably, this threshold field $B_{co}$ is found to be almost temperature independent for $T$ below 40 K (Fig. S6 a). The second feature predicted by the theoritical phase diagram is that at sufficiently high fields, superconducting fluctuations should become negligible and have no significant effect on the competing order. This is indicated by the vertical behaviour of the phase boundary of the competing order (shown by the CM segment in the *B-t* theoritical phase diagram) above the mean critical field $B_{c2}$. This feature is also found in our experimental phase diagram. Indeed, for B ≥ 30 T, the transition line of charge order is almost vertical, showing no significant field effect (this field-independent behaviour of $T_0$ was already reported in ref. 15). This means that what causes the emergence of an electron-like signal in the Hall coefficient, namely Fermi surface reconstruction due to the onset of charge order, becomes field-independent above 30 T or so.

From this comparison, we conclude that our experimental phase diagram bears the fundamental features of a competing order with superconductivity.

**Figure S6 | Phase diagram of competing order: experiment vs theory.**

Comparison of a) the experimental *B-T* phase diagram from high field measurements with b) the *T* = 0 theoritical phase diagram. a) Reproduction of Fig. 2. b) Theoritical *T* = 0 phase diagram of a charge order competing with superconductivity (adapted from ref. 24). SC stands for superconductivity and *t* is a parameter that controls the strength of the competing order. The system is more easily driven to charge order at lower *t*.

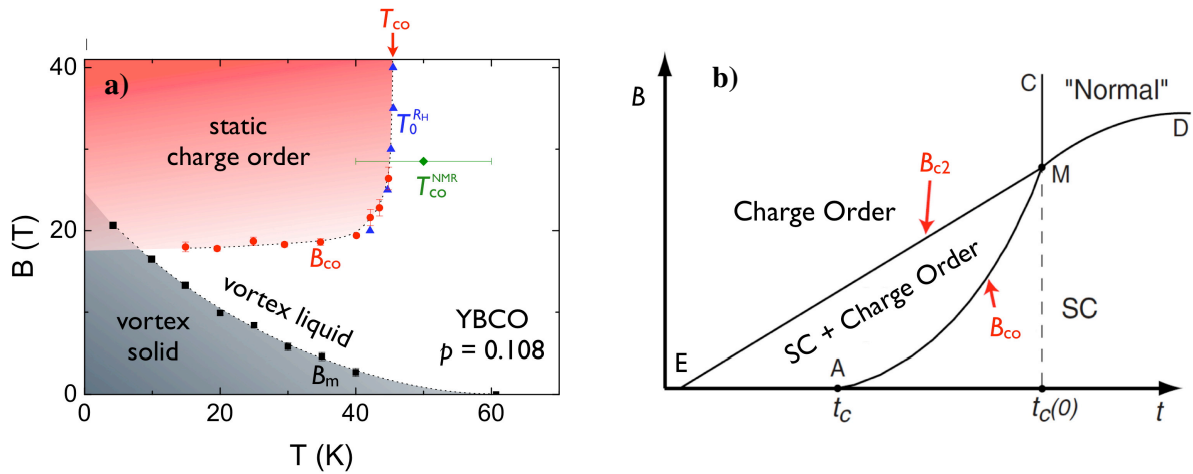

30.     Liang, R. *et al*. Preparation and characterization of highly ordered ortho-II phase in YBa$_2$Cu$_3$O$_{6.50}$ single crystals. *Physica C* **336,** 57 (2000).

31.     Blatter, G. *et al*. Vortices in high-temperature superconductors. *Rev. Mod. Phys.* **66**, 1125 (1994).

32.     Hanaguri, T. *et al*. Elastic properties and anisotropic pinning of the flux-line lattice in single-crystalline La$_{1.85}$Sr$_{0.15}$CuO$_4$. *Phys. Rev. B* **48**, 9772 (1993)

33.     Lei, M. *et al*. Elastic constants of a monocrystal of superconducting YBa$_2$Cu$_3$O$_{7-\delta}$. *Phys. Rev. B* **47**, 6154 (1993)

34.     Fukase, T. *et al.* Ultrasonic studies of anisotropic flux pinning in La$_{1.85}$Sr$_{0.15}$CuO$_4$ under high magnetic fields. *Physica B* **216**, 274 (1996)

35.     Harris, R. *et al*. Phenomenology of a-axis and b-axis charge dynamics from microwave spectroscopy of highly ordered YBa$_2$Cu$_3$O$_{6.50}$ and YBa$_2$Cu$_3$O$_{6.993}$. *Phys. Rev. B* **74**, 104508 (2006).